# Large Anomalous Hall Effect in Topological Insulators with Proximitized Ferromagnetic Insulators

Masataka Mogi[1,*], Taro Nakajima[2], Victor Ukleev[2,3], Atsushi Tsukazaki[4], Ryutaro Yoshimi[2], Minoru Kawamura[2], Kei S. Takahashi[2,5], Takayasu Hanashima[6], Kazuhisa Kakurai[2,6], Taka-hisa Arima[2,7], Masashi Kawasaki[1,2] and Yoshinori Tokura[1,2,8,†]

[1]*Department of Applied Physics and Quantum Phase Electronics Center (QPEC), University of Tokyo, Bunkyo-ku, Tokyo 113-8656, Japan.*

[2]*RIKEN Center for Emergent Matter Science (CEMS), Wako, Saitama 351-0198, Japan.*

[3]*Laboratory for Neutron Scattering and Imaging (LNS), Paul Scherrer Institute (PSI), CH-5232, Villigen, Switzerland.*

[4]*Institute for Materials Research, Tohoku University, Sendai, Miyagi 980-8577, Japan.*

[5]*PRESTO, Japan Science and Technology Agency (JST), Chiyoda-ku, Tokyo 102-0075, Japan.*

[6]*Comprehensive Research Organization for Science and Society (CROSS), Tokai, Ibaraki, 319-1106, Japan.*

[7]*Department of Advanced Materials Science, University of Tokyo, Kashiwa, Chiba 277-8561, Japan.*

[8]*Tokyo College, University of Tokyo, Bunkyo-ku, Tokyo 113-8656, Japan.*

## Abstract

We report a proximity-driven large anomalous Hall effect in all-telluride heterostructures consisting of ferromagnetic insulator $Cr_2Ge_2Te_6$ and topological insulator $(Bi,Sb)_2Te_3$. Despite small magnetization in the $(Bi,Sb)_2Te_3$ layer, the anomalous Hall conductivity reaches a large value of $0.2e^2/h$ in accord with a ferromagnetic response of the $Cr_2Ge_2Te_6$. The results show that the exchange coupling between the surface state of the topological insulator and the proximitized $Cr_2Ge_2Te_6$ layer is effective and strong enough to open the sizable exchange gap in the surface state.

# Main text

In magnetically doped three-dimensional (3D) topological insulator (TI) films, exotic magnetic quantum phases such as a quantum anomalous Hall (QAH) insulator and an axion insulator have been achieved [1-11]. The formation of an exchange gap at the Dirac surface states of TI films and the Fermi-level tuning into the gap are two requisites for the emergence of the topological phenomena, e.g. the chiral edge conduction in the QAH state. In the magnetically doped TI, the size of the energy gap (~50 meV [12]) is produced by the interaction between magnetic impurities and the surface states, whereas it suffers from disorders due to spatial inhomogeneity of magnetic dopants [12,13] and electronic potentials [14]. In fact, the observable temperature of the QAH effect reported so far is lower than about 100 mK in the uniformly Cr- or V-doped $(Bi,Sb)_2Te_3$ films [5,6]. The modulation doping or co-doping technique of the magnetic ions has been developed to reduce the disorder, yet the observable QAH temperature still remains at most around 2 K [7,8]. The ferromagnetic proximity effect is anticipated to be an alternative ideal approach to introduce the uniform magnetic interaction to the surface states [1-3]. The choice of materials for ferromagnetic insulators (FMIs) is a key issue to induce the effective coupling with less disorder; candidates for the FMIs facing the TI is of great variety. Indeed, several FMI/TI heterostructures have been proposed theoretically and synthesized to date [15-29]. Although these studies have demonstrated several potential magnetoelectronic responses, such as magnetoresistance, anomalous Hall effect [17-24], and unconventional surface magnetization [25,26] even at room temperature, the magnitude of the response or the coupling strength to the surface state of TI remains far smaller than expected.

In this Letter, we report a large anomalous Hall effect, being reminiscent of an incipient QAH state, in a FMI/TI heterostructure consisting of $Cr_2Ge_2Te_6$ (CGT) [21,30-33] and $(Bi,Sb)_2Te_3$ (BST) [5-13,22,23,26]. The observation indicates the formation of a sizable exchange gap in the surface state of the TI. Through combined characterizations of the interfacial magnetic property by spin-polarized neutron reflectometry and by magneto-transport measurements, we demonstrate that the exchange coupling is induced by the magnetic proximity effect.

We fabricated the CGT/BST heterostructures on InP(111) substrates by molecular-beam epitaxy (MBE) (see the Supplemental Material [34] for the detailed methods). Ferromagnetic CGT thin layers have recently been achieved not only by mechanical exfoliation of bulk crystals [32] but also by thin film growth with MBE [33]. CGT has a rhombohedral crystal structure of a van der Waals (vdW) type [30], which is presumably matched to the interface formation with a similar triangular lattice constant of BST [Fig. 1(b)] [21,31]. Furthermore, it has been reported that the MBE-grown CGT films possess a perpendicular anisotropic remanent magnetization [33,34], which is favorable to produce the exchange gap in the surface state of TI. Experimentally, the structural characterization of the interface was carried out by a cross-sectional scanning transmission electron microscopy (STEM). Figure 1(c) displays the STEM image of a MBE-grown CGT(12 nm)/BST(9 nm)/CGT(12 nm) heterostructure, exhibiting abrupt interfaces with the ordered stacking orientation in favor of the hexagonal Te arrangements of CGT and BST. By performing Fourier transformation in the lateral direction of the image, the lattice distance of each layer is achieved as depicted in the right panel of Fig. 1(c). Sharp changes of the lateral lattice distance at the interfaces reflect weak epitaxial strain at the interfaces which are a notable feature of the vdW heterointerface. Furthermore, energy dispersive x-ray spectroscopy (EDX) ensures almost no inter-diffusion of atoms [34].

On the basis of the MBE-grown clean heterostructures, we examine the interfacial magnetism of the CGT/BST/CGT sandwiched heterostructure by depth-sensitive polarized neutron reflectometry (PNR) [34]. The PNR measurements, being quantitatively responsive to in-plane magnetization, were conducted at 3 K with an in-plane magnetic field $\mu_0 H_{//} = 1$ T [Fig. 2(a)] which is strong enough to fully align the magnetic moments to the field direction as confirmed by the magnetization hysteresis loops measured at 2 K [Fig. 2(b)]. Figure 2(c) shows the x-ray and non-spin-flip specular PNR reflectivity curves $R^+$ and $R^-$, where $+$ ($-$) denotes the incident neutron spins parallel (antiparallel) to the direction of $H_{//}$ as a function of the momentum transfer vector $Q_z$. The in-plane saturated magnetization of the sample is directly reflected in the spin-asymmetry ratio defined as $(R^+ - R^-)/(R^+ + R^-)$ [Fig. 2(d)]. In addition, we combined an x-ray reflectivity (XRR) measurement at room temperature [Fig. 2(c)] to conduct a model analysis for the structural parameters including thickness and roughness of each layer. The depth profile of the x-ray scattering length density (SLD) shown in Fig. 2(e), corresponding to the electron density distribution in the heterostructure, reflects the structural interface roughness. Notably, the root mean square roughness of all interfaces in the SLD profiles is less than 1 nm, which is consistent with the STEM image shown in Fig. 1(c). The structural parameters derived from the XRR fitted model were used to refine the PNR curves. Figure 2e displays the magnetic SLD depth-profiles based on the fitting results on the $R^+$, $R^-$ [Fig. 2(c)] and the spin-asymmetry ratio [Fig. 2(d)], taking into account the structural depth profile obtained from the XRR data. The fitting analysis yields magnetizations of 152±8 emu/cm$^3$ and 0±20 emu/cm$^3$ for the CGT and BST layers [34], respectively. Although it is difficult to precisely determine the induced magnetization in the BST layer due to the spatial broadening, it will be reasonable to conclude from the present fitting analysis that the induced magnetization in the BST layer is far smaller than the magnetization of the CGT layer.

The ferromagnetic proximity effect on the surface states of the TI can be assessed by magneto-transport measurements. The measurements were conducted with the sandwiched CGT/BST/CGT trilayers and the CGT/BST bilayers. Because of the high electric resistance of the CGT layer [33], its contribution to electrical transport is negligibly small [34]. For the TI layer, instead of simple single-layered $(Bi_{1-x}Sb_x)_2Te_3$, we engineered a multi-layer structure of $(Bi_{1-x}Sb_x)_2Te_3$(2 nm)/$Bi_2Te_3$(2 nm)/$(Bi_{1-x}Sb_x)_2Te_3$(2 nm) [Fig. 3(a)] which works as a conduction channel. The reason for adopting the multi-layer structure is as follows. In the CGT/BST/CGT heterostructures, the charge neutrality point takes place at a relatively small value of $x$ ($0.3 < x < 0.4$) due to possible hole transfer from CGT to BST [34]. According to an ARPES study on BST [41], small $x$ causes the Dirac point to submerge below the bulk valence band. To approach the Dirac point with the charge neutrality condition, we need to dope electrons while keeping $x > 0.5$. To fulfill this requirement, we inserted the electron-rich $Bi_2Te_3$ layer between the BST layers to assist electron doping. The value of $x$ in the BST layer is kept larger than 0.5 assuming that the surface band structure is mainly affected by the environment near the interface [36]. Consequently, we could prepare the samples with low carrier density at reasonably large Sb compositions, $x = 0.6$ and 0.64, which show semiconducting temperature ($T$) dependence of the longitudinal sheet resistivity ($\rho_{xx}$) as depicted in Fig 3(b).

In these samples, large anomalous Hall resistance (> 1 k$\Omega$) appear with perpendicular anisotropic hysteresis loops as shown in Fig. 3(c). We show in Fig. 3(d) the $x$ dependence of the sheet carrier density ($n_{2D}$), the longitudinal sheet conductivity ($\sigma_{xx}$) and the Hall conductivity ($\sigma_{xy}$) at zero fields as converted from $\rho_{xx}$ and $\rho_{yx}$. The notable feature is that the $\sigma_{xy}$ exceeds $0.2e^2/h$ in the most insulating sample ($x = 0.6$) where $\sigma_{xx} \sim 2e^2/h$ and $n_{2D} \sim 10^{12}$ cm$^{-2}$ [46]. The sheet carrier densities are estimated from the slope of Hall resistance above the saturation field. The carrier types are electrons for $x = 0.3$ and holes for $x = 0.64$, demonstrating that the Fermi level is systematically shifted from $n$-type to $p$-type with increasing $x$ [Fig. 3(c)].

These observations in Fig. 3 can be understood by the opening of an exchange gap in the dispersion relation of the TI surface state. When an exchange gap opens on the surface of TI, the Berry curvature is strongly enhanced near the band edge resulting in the large $\sigma_{xy}$. At the same time, when the Fermi energy is tuned within or close to the exchange gap, $\sigma_{xy}$ takes a maximum while $\sigma_{xx}$ takes a minimum. In the present study, we observe that $\sigma_{xy}$ takes a maximum accompanied by a nearly minimum value of $\sigma_{xx}$ in the sample with the low carrier density ($x = 0.6$) as expected. Also, the decrease in $\sigma_{xy}$ and increase in $\sigma_{xx}$ are observed as the carrier density is detuned from the optimum value. These carrier density dependencies are consistent with the picture described above, indicating the opening of the exchange gap by the magnetic proximity effect. The increase in $\sigma_{xy}$ accompanied by the decrease in $\sigma_{xx}$ leads to an enhancement in the Hall angle $\theta_H = \tan^{-1}(\sigma_{xy}/\sigma_{xx})$, discriminating the anomalous Hall effect of extrinsic origin [45]. The obtained values of the $\sigma_{xy}$ and the $\theta_H$ are dramatically increased in our samples compared to those reported in other FMI/TI systems [Fig. 3(e)] [34]. This trend suggests that the Fermi level is close to the exchange gap and/or that the exchange gap is large in our samples compared among those FMI/TI systems, although the quantitative estimation of the size of the exchange gap is difficult due to residual disorder/inhomogeneity in the samples [43,44] (see [34] for the detailed discussion).

The CGT-layer thickness dependence provides additional evidence that the observed anomalous Hall effect is induced by the magnetic proximity effect, excluding other origins arising from Cr diffusion into the BST layer. Figure 4(a) shows the temperature dependent magnetization (*M-T*) curves of four CGT/BST bilayers [the inset of Fig. 4(c)] films with representative CGT-layer thicknesses under $\mu_0 H_\perp = 0.05$ T. Both magnetization $M$ and Curie temperature $T_C$ decrease systematically with decreasing the CGT film thickness $t$. As shown in Fig. 4(b), the low-temperature values of $\sigma_{xy}$ also decrease with decreasing $t$. In Fig. 4(c), the $t$ dependencies of $\sigma_{xy}$ at $T = 2$ K and the saturated magnetization of CGT $M_s$ are plotted together. The agreement in $t$-dependencies of $\sigma_{xy}$ and $M_s$ indicates that $\sigma_{xy}$ is almost proportional to $M_s$. This observation shows that the exchange gap on the surface state of the TI can be tuned by the magnetization of the CGT layer, directly pointing to the proximity-coupling origin of the anomalous Hall effect. The decrease in $M_s$ in the range of $t < 2$ nm is attributed perhaps to the finite size effect of the 2D ferromagnetic CGT layer [31]. In contrast to $\sigma_{xy}$, $\sigma_{xx}$ is almost constant with variation of $t$ across $t \sim 2$ nm [Fig. 4 (c)]. The constancy of $\sigma_{xx}$ suggests that the Fermi energy and the scattering time are not largely affected by the thickness of the CGT layer.

On the basis of the above experimental results, we discuss possible mechanisms of the exchange gap formation at the CGT/BST interface. One conceivable scenario would be the induction of magnetization in the TI layer by the adjacent FMI layer as discussed in the earlier works [15-17,25,26,28]. However, this scenario is unlikely applicable to the present case. At the interfaces of EuS/Bi$_2$Se$_3$ and EuS/BST, large magnetizations of about 270 and 160 emu/cm$^3$, respectively, in the TI layer have been reported [25,26]. In contrast, in the present study, the magnetization in the CGT layer is already smaller than these values. Therefore, the induced magnetization, even if it existed, in the BST layer of the present CGT/BST heterostructure would be much smaller than that reported for the EuS-based heterostructures [25,26]. Despite the small induced magnetization, our transport measurements have revealed that the $\sigma_{xy}$ and $\theta_H$ are much enhanced in the CGT/BST system. One other possible scenario to understand these observations is the formation of the exchange gap by penetration of the TI surface state wave function into the FMI layer. In this scenario, even if the interfacial magnetization in the BST layer is small, the penetrated part of the surface state wave function can interact with the magnetic moment in the CGT to produce a sizable exchange gap. A recent first-principle calculation work [29] indicates the formation of a large exchange gap in Te-based heterostructure $MnBi_2Te_4/Bi_2Te_3$ based on the wave function penetration mechanism.

In summary, we have synthesized CGT/BST/CGT heterostructures and have studied the ferromagnetic proximity effect at the interface of the heterostructures. We have observed the depth profile of the magnetization by the PNR measurement which suggests small induced magnetization in the BST layer. We have also observed a large anomalous Hall angle in magneto-transport measurements, which indicates a sizable exchange gap. To explain both observations, we have proposed the exchange gap formation due to the penetration of the TI surface state wave function into the FMI layer.

We acknowledge helpful discussions with K. Yasuda, R. Watanabe, R. Fujimura, Y. Fujishiro, Y. Okamura, Y. Kaneko, S. Maekawa, J. G. Checkelsky and N. Nagaosa. This research was partly supported by JSPS/MEXT Grant-in-Aid for Scientific Research (No. 15H05853, 15H05867, 17J03179, 18H04229, 18H01155), and JST CREST (No. JPMJCR16F1). This neutron experiment at MLF, J-PARC was performed under a user program (Proposal No.2017B0195).*mogi@cmr.t.u-tokyo.ac.jp
†tokura@riken.jp

Figures

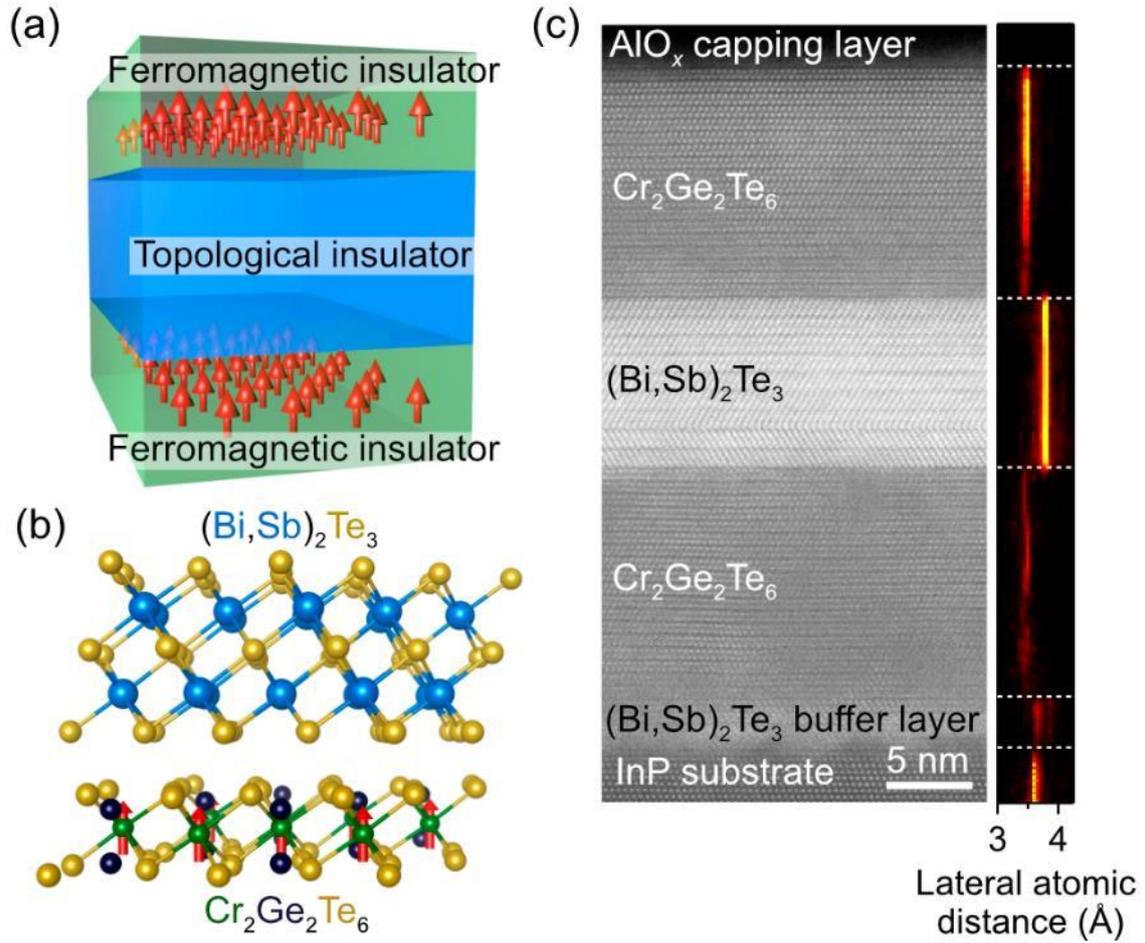

FIG. 1. (a) Schematic drawing of the interfacial exchange coupling in a 3D topological insulator (TI) sandwiched by ferromagnetic insulators (FMIs). (b) Schematic of crystal structures of TI $(Bi,Sb)_2Te_3$ (BST) and FMI $Cr_2Ge_2Te_6$ (CGT) with the relationship of stacking orientation expected from their Te arrangements in the respective layer planes. (c) Cross-sectional high-angle annular dark-field STEM image of the CGT(12 nm)/BST(9 nm)/CGT(12 nm) heterostructure grown on an InP substrate with a BST buffer layer (left panel). The lateral atomic distance of each layer obtained by the Fourier transformation of the left image plotted along the growth direction (right panel).

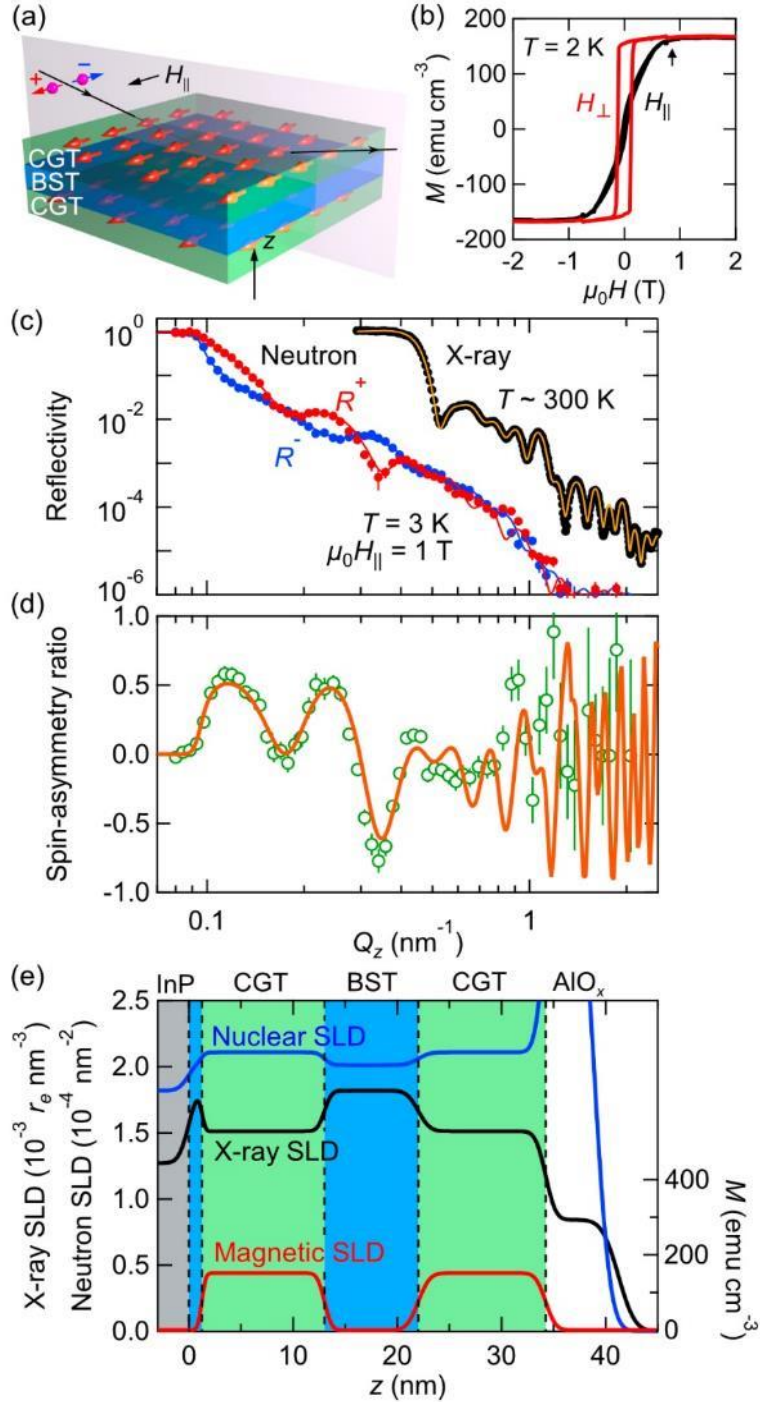

FIG. 2. (a) Schematic of the PNR experimental set-up for the CGT/BST/CGT structure. (b) Magnetization hysteresis loops under out-of-plane ($H_\perp$) and in-plane ($H_\parallel$) magnetic fields for the identical CGT(12 nm)/BST(9 nm)/CGT(12 nm) sample used in the PNR experiments. The black arrow represents the saturation field for in-plane direction. (c) Measured (dots) and fitted (solid lines) reflectivity curves for the x-ray (black), neutron of spin-up ($R^+$) (red) and spin-down ($R^-$) (blue) as a function of momentum transfer ($Q_z$) on a logarithmic scale. The error bars represent one standard deviation. (d) PNR spin-asymmetry ratio ($R^+ - R^-$)/($R^+ + R^-$) obtained from experimental and fitted reflectivity curves in (c). The error bars represent one standard deviation. (e) X-ray scattering length density (SLD) (black), and neutron SLD divided into the nuclear (blue) and the magnetic (red) SLDs as a function of the distance from the InP substrate surface ($z$). The $r_e$ in the unit of the X-ray SLD denotes the classical electron radius of 2.8179...×10$^{-15}$ m. For the magnetic SLD, the value of $M$ corresponding to the neutron SLD is shown in the right ordinate.

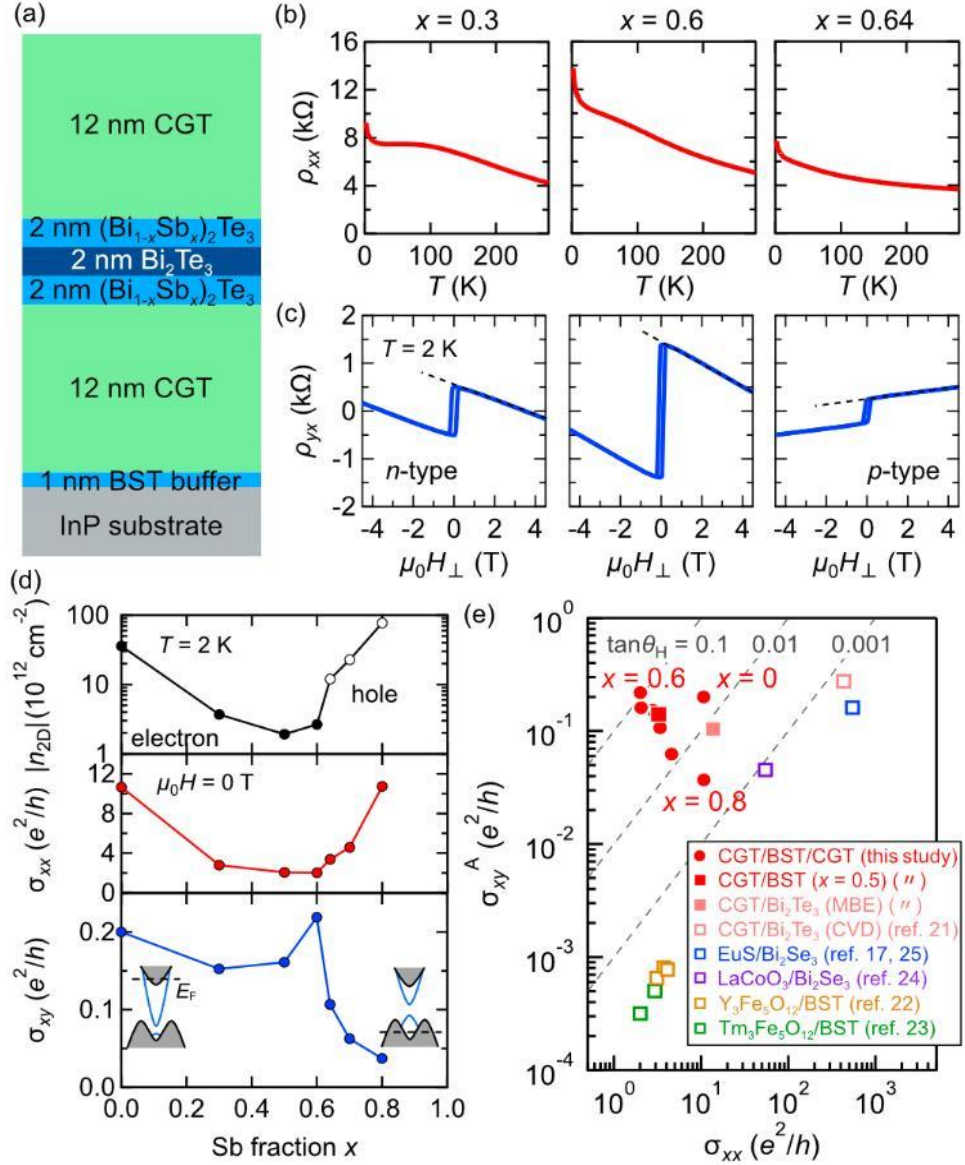

FIG. 3. (a) Schematic layout of CGT(12 nm)/$(Bi_{1-x}Sb_x)_2Te_3$(2 nm)/$Bi_2Te_3$(2 nm)/$(Bi_{1-x}Sb_x)_2Te_3$(2 nm)/CGT(12 nm) heterostructure. (b, c), Temperature ($T$) (out-of-plane magnetic field ($\mu_0 H_\perp$)) dependence of the longitudinal sheet resistivity ($\rho_{xx}$) in zero-field (b) (the Hall resistivity ($\rho_{yx}$) at 2 K (c)) of CGT/$(Bi_{1-x}Sb_x)_2Te_3$/$Bi_2Te_3$/$(Bi_{1-x}Sb_x)_2Te_3$/CGT ($x$ = 0.3, 0.6, 0.64) heterostructures. (d) Sb fraction ($x$) dependence of the sheet carrier density ($|n_{2D}|$) (top panel), the longitudinal sheet conductivity ($\sigma_{xx}$) (middle panel), and the Hall conductivity ($\sigma_{xy}$) (bottom panel) at 2 K. The insets are the simplified schematics of band structures representing the different Fermi energies; the blue lines represent the dispersion of the surface state. (e) The anomalous Hall conductivity $\sigma_{xy}^A$ is plotted against $\sigma_{xx}$ plot with the use of the data for the present heterostructures shown in (d) in comparison with other various FMI/TI heterostructures [17, 21-25]. The values of $\sigma_{xy}^A$ and $\sigma_{xx}$ are taken from the data obtained at the lowest temperature in the measurements (2-5 K) for the magnetized state where the out-of-plane magnetization saturates.

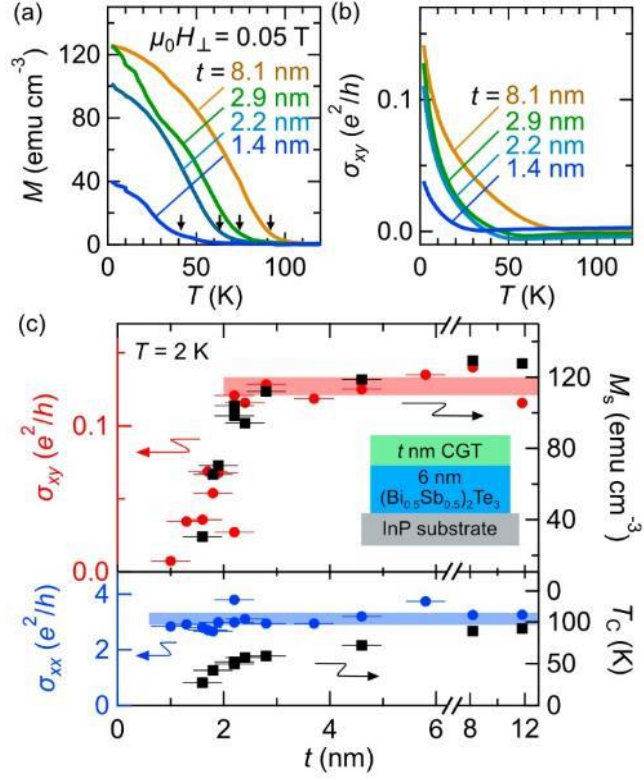

FIG. 4. (a, b) M-T (a) and $\sigma_{xy}$-T (b) curves measured under field cooling with $\mu_0 H_\perp = 0.05$ T in BST/CGT ($t$ = 1.4, 2.2, 2.9 and 8.1 nm) bilayer structures. Black arrows indicate the Curie temperature $T_C$ for highlighting the changes against $t$. (c) CGT thickness $t$ dependence of $\sigma_{xy}$ (red closed circles, left red ordinate) and the spontaneous magnetization $M_s$ (black closed squares, right black ordinate) at 2 K under zero magnetic field (top panel), and $\sigma_{xx}$ at 2 K (blued closed circles, left blue ordinate) and $T_C$ (black closed squares, right black ordinate) (bottom panel). The inset shows the schematic layout of $(Bi_{0.5}Sb_{0.5})_2Te_3$ (6 nm)/CGT ($t$ nm) bilayer structure.

Supplementary Material for

# Large Anomalous Hall Effect in Topological Insulators with Proximitized Ferromagnetic Insulators

M. Mogi, T. Nakajima, V. Ukleev, A. Tsukazaki, R. Yoshimi, M. Kawamura,

K. S. Takahashi, T. Hanashima, K. Kakurai, T. Arima, M. Kawasaki and Y. Tokura

**I. Methods.**

**Sample fabrication.**

The CGT/BST heterostructures were grown on semi-insulating ($> 10^7$ Ω cm) InP(111)A substrates using standard Knudsen cells in a MBE chamber under a vacuum condition ($\sim 1 \times 10^{-7}$ Pa). The respective layers were grown in the same procedures as described in Refs. [S1,S2]. Before the growth of the CGT, 1-2 quintuple layers of BST were grown as an insulating buffer layer to improve the crystallinity of the CGT. For the Fermi-level tuning in the trilayers depicted in Fig. 3, we modulated Bi/Sb ratio by open and close of the shutter for the Sb cell and succeeded in the tuning as investigated by Chang *et al.* [S3]. Taking the films out of the MBE chamber, the AlO$_x$ capping layer (~5 nm) was immediately deposited by an atomic layer deposition (ALD) system at room temperature to prevent deterioration of the films. Thicknesses of respective layers were determined by XRR measurements. For transport measurements, the films were patterned into Hall bars with 200-300 μm in width and 700-1,000 μm in length by using photolithography and chemical wet etching by H$_2$O$_2$/H$_3$PO$_4$/H$_2$O (1:1:8) and HCl-H$_2$O (1:4) mixtures. Electrical contact was made of Ti(3 nm)/Au(27 nm) by electron beam evaporation.

**Polarized neutron and x-ray reflectometry.**

PNR experiments were performed at BL-17 SHARAKU of the Materials and Life Science Experimental Facility (MLF) in the Japan Proton Accelerator Research Complex (J-PARC), Tokai,

Japan [S4,S5]. The sample was loaded into a closed cycle refrigerator and was cooled to 3 K. An electromagnet was used to apply a magnetic field of 1 T along the in-plane direction of the film, which was perpendicular to the incident neutron beam and neutron momentum transfer, $Q_z$. The PNR spectra were measured by means of the time-of-flight technique with a pulsed polychromatic incident neutron beam; the wavelength range was from 2.4 to 8.8 Å. We selected three different incident angles (0.3, 0.9 and 2.7 degrees) to provide access to the momentum transfer range from 0.08 to 2 nm$^{-1}$. A supermirror polarizer, guide-field coils, and a spin flipper were employed to obtain the polarized incident neutron beam, whose polarization direction was set to be parallel or antiparallel to the external magnetic field at the sample position. The beam polarization was approximately 98 %. The intensities of reflected neutrons were measured without analyzing the spin state of the neutrons and converted to the PNR spectra by the UTSUSEMI software [S6], in which effects of the imperfect beam polarization were corrected. The PNR spectra were fitted using GenX software [S7], assuming that the magnetic moments of Cr were parallel to the magnetic field. Complementary XRR spectrum was measured by Bruker D8 x-ray diffractometer at room temperature to determine layer thicknesses of the sample. The sample used in the XRR measurement is identical to that used in the PNR measurements. An incident x-ray beam with a wavelength of 1.5406 Å was obtained by Cu Kα radiation. The intensity of specular reflection was measured with varying the incident angle from 0.4 to 4.3 deg to cover $Q_z$ range from 0.3 to 3 nm$^{-1}$.

**Transport and magnetization measurements.**

Electrical transport and magnetization measurements were carried out using a Quantum Design physical property measurement system (PPMS) and a magnetic property measurement system (MPMS) superconducting quantum interference device (SQUID) magnetometer, respectively, in the temperature range from 2 to 300 K. $\sigma_{xy}$ and $\sigma_{xx}$ in the four terminal measurements were converted from $\rho_{yx}$ and $\rho_{xx}$ following the tensor relations $\sigma_{xy} = \rho_{yx}/(\rho_{xx}^2+\rho_{yx}^2)$ and $\sigma_{xx} = \rho_{xx}/(\rho_{xx}^2+\rho_{yx}^2)$.

## II. Characterization and fundamental properties of CGT/BST/CGT heterostructures.

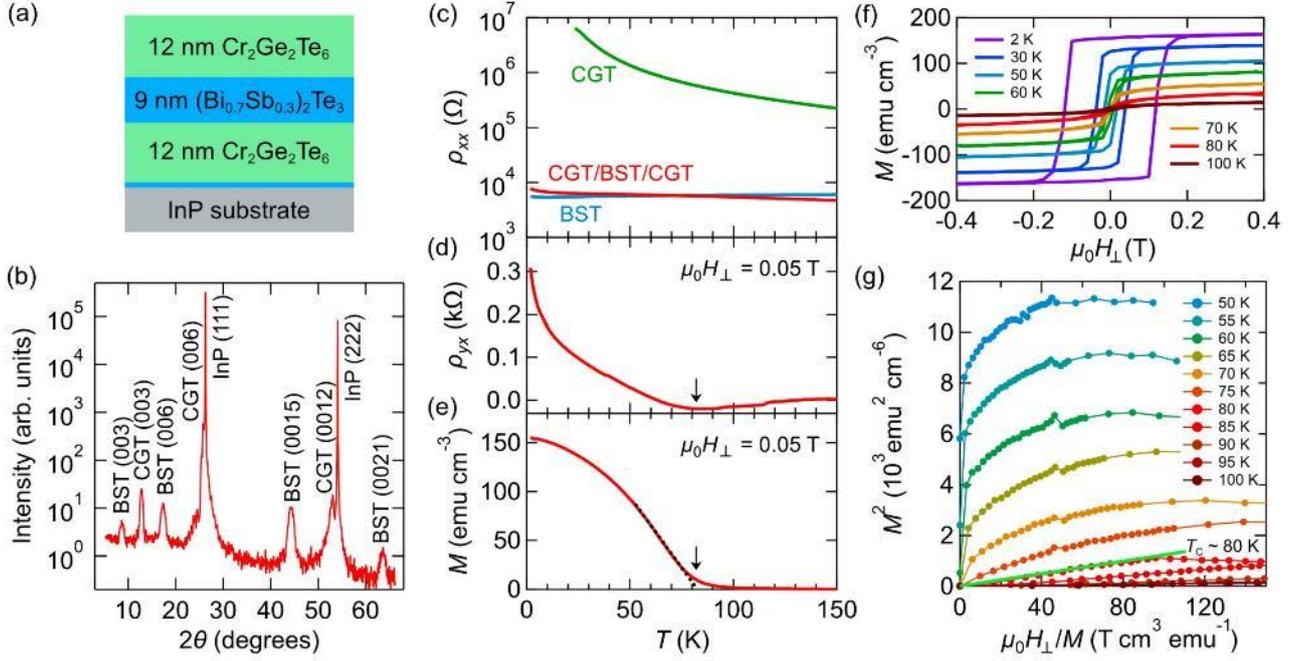

FIG. S1. Characterization of the CGT/BST/CGT heterostructure prepared for STEM and PNR. (a) Cross-sectional schematic of the CGT(12 nm)/BST(9 nm)/CGT(12 nm) on a 1-nm-thick BST buffered InP substrate, where the Sb fraction of BST is $x = 0.3$, used for STEM/EDX and PNR studies in the main text. (b) XRD pattern on a logarithmic scale for the CGT/BST/CGT structure shown in (a). (c-e) Temperature ($T$) dependence of the longitudinal sheet resistivity ($\rho_{xx}$) (c), the Hall resistivity ($\rho_{yx}$) (d) and the magnetization ($M$) (e) of the CGT/BST/CGT structure shown in (a). In (c), $\rho_{xx}$ of a CGT single-layer ($t = 12$ nm) grown on a 1-nm-thick BST buffered InP substrate and a BST single-layer ($t = 9$ nm) directly grown on an InP substrate are also shown. Black arrows indicate the rising temperature of $\rho_{yx}$ and $M$. (f) Perpendicular magnetic field ($\mu_0 H_\perp$) dependence of the magnetization ($M$) at various temperatures. (g) Arrott plot to determine the Curie temperature ($T_C \sim 80$ K).

Figure S1 shows the x-ray diffraction (XRD) characterization and fundamental physical properties of the CGT/BST/CGT sandwiched heterostructure [Fig. S1(a)], which was characterized by STEM and PNR in the main text. Figure S1(b) shows the XRD pattern for the heterostructure. We observe diffraction peaks from both BST and CGT layers as expected. Figure S1(c) shows the electrical resistivity ($\rho_{xx}$) of the heterostructure. The $\rho_{xx}$ of a CGT single-layer ($t = 12$ nm) is more than two orders of magnitude larger than that of a BST single-layer ($t = 9$ nm), ensuring the least contribution of parasitic conduction in the CGT layer. In fact, $\rho_{xx}$ of CGT/BST/CGT structure is comparable to that of a 9-nm-thick BST single-layer. Figure S1(d) shows the temperature ($T$) dependence of the Hall resistivity ($\rho_{yx}$). With decreasing $T$, $\rho_{yx}$ and $M$ rise at around $T = 80$ K as highlighted by black arrows in Fig. S1(e). In Figs. S1(f) and (g), we show the magnetic hysteresis loops at various temperatures and the Arrott plot, respectively. We find the Curie temperature $T_C \sim 80$ K in accord with the $M$-$T$ curve [Fig. S1 (e)]. It should be noted that the $T_C$ is slightly higher than that of the bulk crystals. In fact, $T_C$ of the MBE-grown CGT films alone also shows a similar enhanced $T_C$ possibly due to some defects in the films [S1].

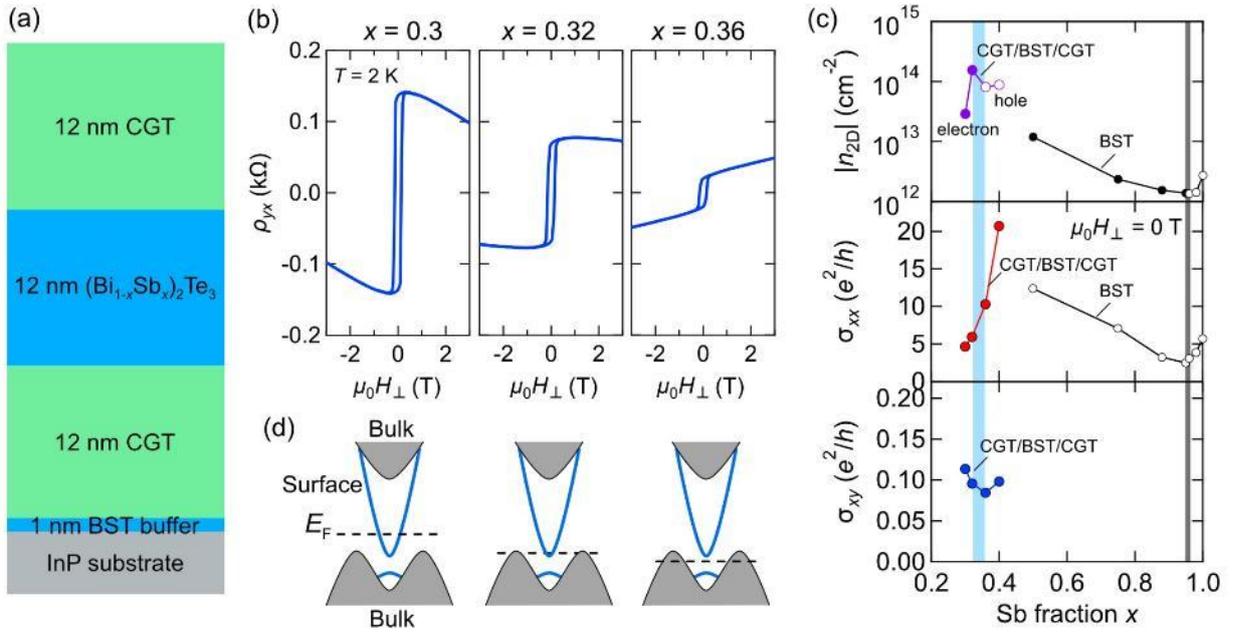

FIG. S2. Transport properties of CGT/BST/CGT sandwiched heterostructures. (a) Cross-sectional schematic of the CGT(12 nm)/BST(12 nm)/CGT(12 nm) on a 1-nm-thick BST buffered InP substrate (b) Out-of-plane magnetic field ($B$) dependence of the Hall resistivity ($\rho_{yx}$) at $T = 2$ K for the CGT/(Bi$_{1-x}$Sb$_x$)$_2$Te$_3$/CGT ($x = 0.3, 0.32, 0.36$) heterostructures. (c) Sb fraction ($x$) dependence of the sheet carrier density ($|n_{2D}|$) (top panel), the longitudinal sheet conductivity ($\sigma_{xx}$) (middle panel), and the Hall conductivity ($\sigma_{xy}$) (bottom panel) in CGT/BST/CGT and BST single-layer films. The data for the BST single-layer films is an excerpt from [S8]. (d) Schematics of the relationship between the Fermi-level ($E_F$) and the TI band structure for the respective films which are shown in (b).

For transport measurements described in the main text (Fig. 3), we used the $(Bi_{1-x}Sb_x)_2Te_3/Bi_2Te_3/(Bi_{1-x}Sb_x)_2Te_3$ heterostructures sandwiched by CGT layers to tune the Fermi-level close to the Dirac point. To explain the effectiveness of this structure, we show measurements for $(Bi_{1-x}Sb_x)_2Te_3$ single-layers sandwiched by CGT [Fig. S2(a)]. The Hall responses are shown in Fig. S2(b) for representative three samples ($x = 0.3, 0.32, 0.36$). One important result here is the discrepancy between the charge neutrality point (between $x = 0.32$ to $0.36$) and the maximal anomalous Hall resistivity composition ($x < 0.32$) as summarized by the Sb fraction ($x$) dependence in Fig. S2(c). We assign the discrepancy to the hole conduction in the bulk region as illustrated in Fig. S2(d), where the Dirac point buries in the bulk valence band as reported in the angle-resolved photoemission study for Bi-rich BST ($x \sim 0.3$) [S8]. In such a situation, even if the Fermi-level is tuned near the Dirac point, parasitic conduction in the bulk valence band appears. Furthermore, the value of $x \sim 0.32$, corresponding to the Fermi-level roughly tuned at the charge neutrality point in the CGT/BST/CGT structures, is much lower than that for pristine BST ($x \sim 0.8$) films [S8]. We preliminarily attribute the shift of the charge neutrality point to a charge transfer between $p$-type CGT and $n$-type BST layers. When holes are transferred to the BST layer, it is necessary to compensate the carriers by introducing electrons with lowering $x$. In fact, the MBE-grown CGT films show the weak $p$-type conduction possibly because of Ge deficiency [S1], which can generate holes.

To overcome this situation, we applied the $(Bi_{1-x}Sb_x)_2Te_3/Bi_2Te_3/(Bi_{1-x}Sb_x)_2Te_3$ heterostructures to suppress the bulk hole conduction with the electron-rich $Bi_2Te_3$ thin layer at the inner region. Such a modulation doping technique is adopted for conventional semiconductor heterostructures [S9]. In addition, the proximitized region should be carefully designed to maximize the Berry curvature at the Dirac point. We insert the $(Bi_{1-x}Sb_x)_2Te_3$ layer for the Fermi-level tuning of the isolated Dirac point in the bulk band gap because the surface band structure is strongly affected by the environment near the surface [S3].

**III. Cross-sectional elemental distribution in the CGT/BST/CGT heterostructure.**

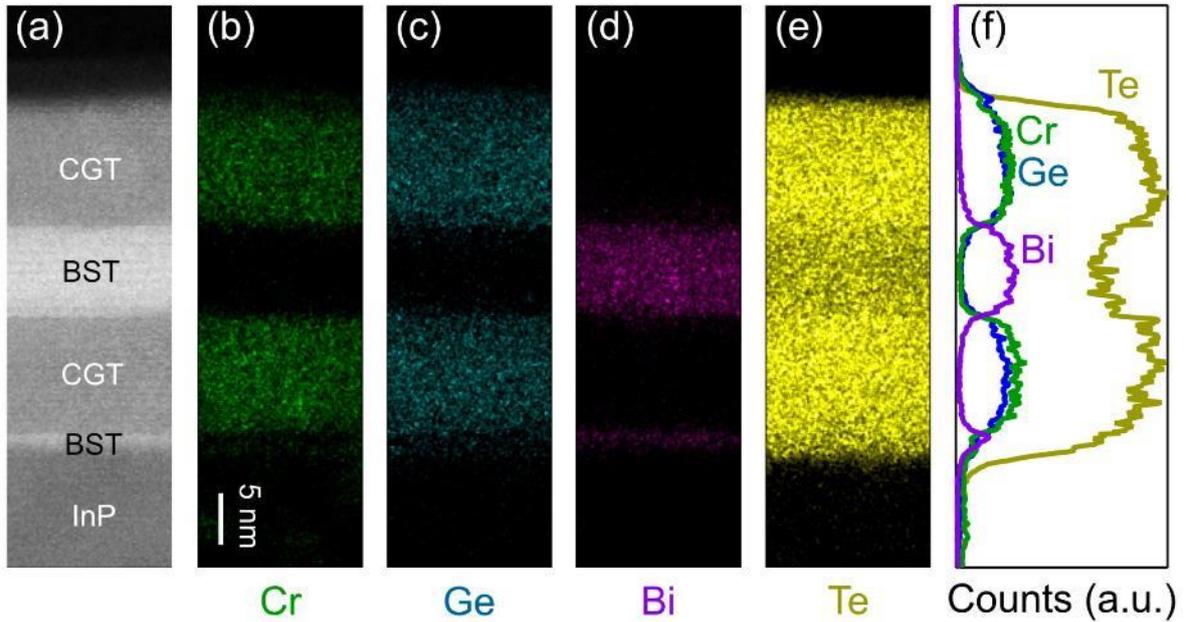

FIG. S3. STEM-EDX for the CGT/BST/CGT heterostructure. (a) STEM image of CGT/BST/BST corresponding to the EDX scan area. (b-e) Distribution maps of each element, Cr (b), Ge (c), Bi (d), and Te (e). (f) Line profiles of Cr, Ge, Bi, and Te.

Figure S3 shows energy dispersive x-ray spectroscopy (EDX) mappings of Cr (b), Ge (c), Bi (d), and Te (e) for the CGT(12 nm)/BST(9 nm)/CGT(12 nm) grown on a 1-2 nm BST buffered InP substrate, which is an identical sample measured by STEM in the main text [Fig. 1(c) and Fig. S3(a)]. Uniform chemical composition is detected as expected for the constituted layer structure. The line profiles in Fig. S3(f) indicate that no discernible inter-diffusion of magnetic Cr atoms is detected.

**IV. Scheme of fitting and alternative models for the PNR reflectivity data.**

Firstly, we explain the scheme of fitting the data shown in Fig. 2 of the main text, in which we show the best-fitted data, indicating that the magnetization resides almost only in the CGT layers. The refinement of the PNR and the XRR fittings was performed using GenX software [S7] for a multilayered model consisting of InP substrate/BST buffer layer/CGT/2 quintuple layer (QL) interfacial BST/BST/2-QL interfacial BST/CGT/AlO$_x$ capping layer. Fitting parameters were layer thicknesses, densities, and interfacial roughness. Furthermore, we introduced magnetization in the CGT and the 2-QL interfacial BST layers as fitting parameters. We note that we use models containing different thicknesses of the AlO$_x$ capping layer for the convergence of XRR and PNR simulations possibly due to a little deterioration of the capping layer in the interval of their experiments. The fitting curves were evaluated by the logarithmic figure of merit taking errors into account (FOM$_{\text{logbar}}$) as below,

$$\text{FOM}_{\text{logbar}} = \frac{1}{N-p} \sum_i \frac{\log_{10} Y_i - \log_{10} S_i}{E_i \log_{10} Y_i},$$

where $N$ is the total number of data consisting of XRR and PNR reflectivity data, $p$ is the number of fitting parameters, $Y_i$ and $S_i$ represent the experimental and simulated dataset, respectively, $E_i$ is the uncertainty in the experimental data, and $i$ indicates individual elements of the dataset. By minimizing FOM$_{\text{logbar}}$ (= 0.99), we obtain the magnetization of 152 emu/cm$^3$ for the CGT layer and of –0.4 emu/cm$^3$ for the interfacial BST layer.

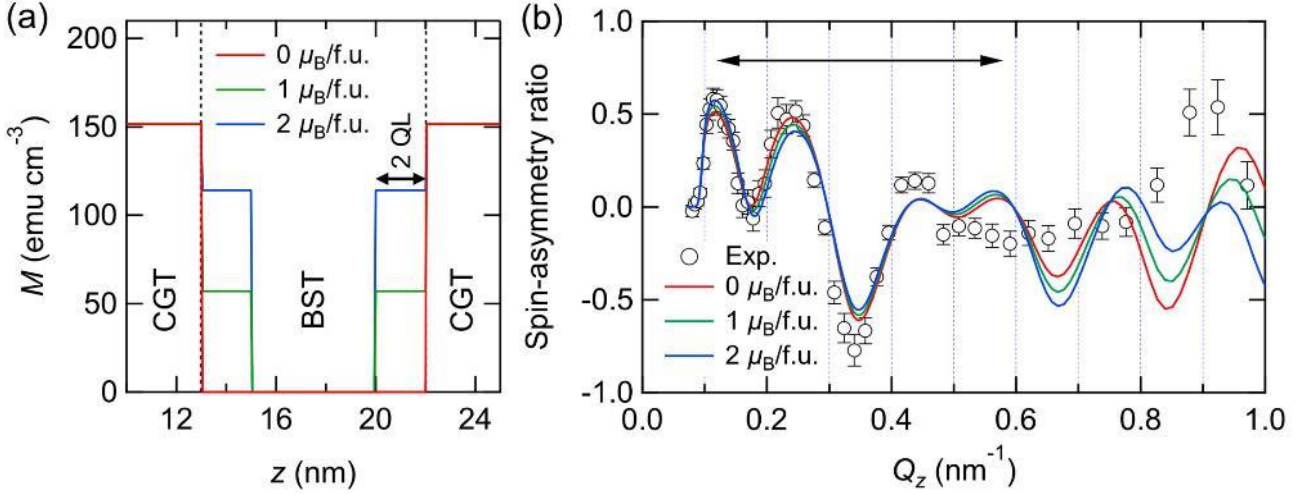

FIG. S4. Fitting curves for additional models. (a) Magnetization depth profiles, in which the interfacial roughness is not included for the clarity, for simulation of CGT/BST/BST structural models with interfacial magnetization in 2 quintuple layers (QL) in BST (0, 1, 2 $\mu_B$ per formula unit of BST). (b) Simulations of PNR spin-asymmetry ratio (solid lines) with experimental data (open dots). The double-headed arrow represents the most statistically reliable region of $Q_z$.

We further investigate the magnetization at the interface (Fig. S4). We prepare models in which we intentionally introduce the magnetization (1 and 2 $\mu_B$/f.u. for BST) at the 2-QL interfacial BST region as shown in Fig. S4(a), following the PNR study on EuS/Bi$_2$Se$_3$ films [S10]. Figure S4(b) displays the simulation data corresponding to the models in Fig. S4(a). The data are well fitted by the model in the range of $0.1 < Q_z < 0.6$ although a clear difference is difficult to see among the results with the fitting parameters $M = 0$, 1, and 2 $\mu_B$/f.u. assumed for the interfacial 2-QL BST.

**V. Additional comparison of anomalous Hall effect with various FMI/TI structures.**

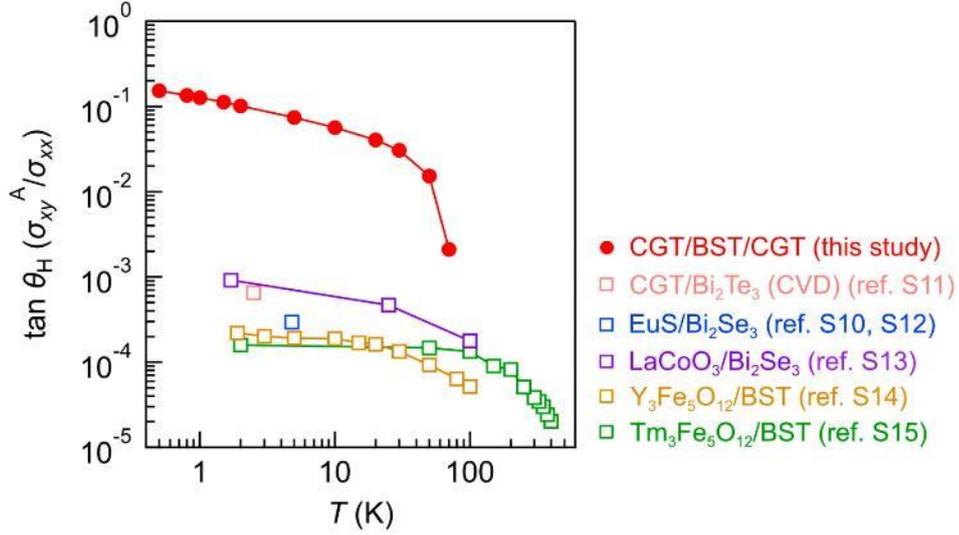

FIG. S5. Tangent anomalous Hall angle ($\sigma_{xy}^A/\sigma_{xx}$) versus temperature ($T$) of CGT/BST and various FMI/TI heterostructures.

In addition to the $\sigma_{xx}$-$\sigma_{xy}^A$ plane plot shown in Fig. 3(e) of the main text, we plot the tangent Hall angle ($\sigma_{xy}^A/\sigma_{xx}$) for the CGT/BST heterostructures compared with previously reported various FMI/TI heterostructures (Fig. S5). The tangent Hall angle is two or three orders of magnitude larger than those of other heterostructures including CGT/chemical vapor deposition (CVD)-grown $Bi_2Te_3$ (2.5 K) [S11], EuS/$Bi_2Se_3$ (< 5 K) [S10,S12] LaCoO$_3$/$Bi_2Se_3$ (1.7-100 K) [S13], $Y_3Fe_5O_{12}$/BST (1.9-200 K) [S14] and $Tm_3Fe_5O_{12}$/BST (2-400 K) [S15]. We note that the present MBE-grown CGT/BST heterostructure was greatly improved as compared with the previously studied CGT/$Bi_2Te_3$ heterostructure grown by chemical vapor deposition (CVD) [S11], possibly because a thick (30 nm) $Bi_2Te_3$ layer with a high bulk carrier density was used in the previous study, on which the ferromagnetic proximity effect would be masked.

**VI. Disorder/inhomogeneity in samples and estimation of the size of the exchange gap.**

We discuss an attempt to estimate the size of the exchange gap from the transport results on the basis of a simple Dirac fermions model, although the estimation is impeded by the presence of disorder/inhomogeneity in samples as below. We firstly describe the influence of disorder/inhomogeneity in our samples from the temperature-dependent transport data.

(i) Influence of disorder/inhomogeneity in the samples

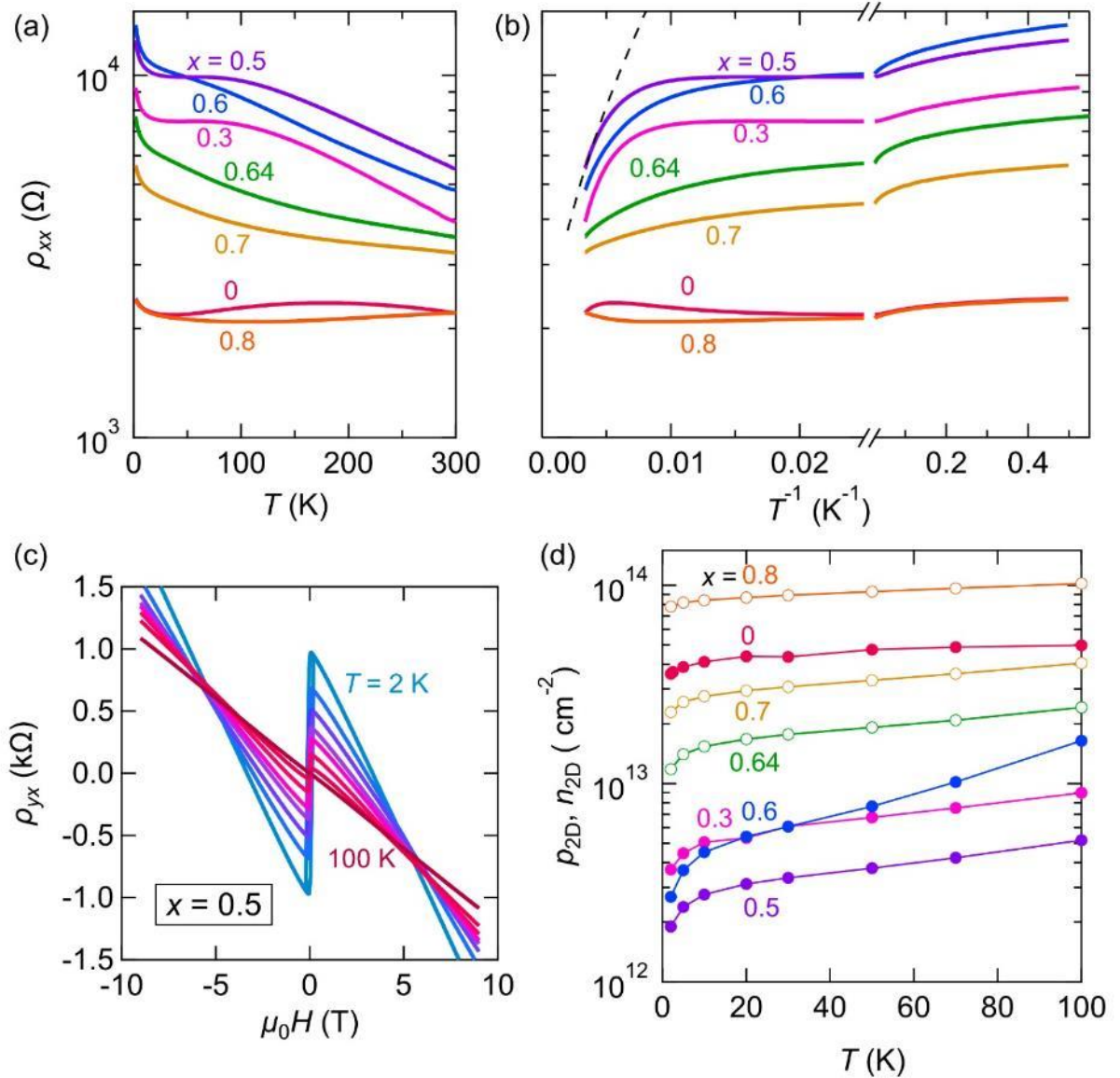

FIG. S6. Temperature-dependent transport properties for CGT(12 nm)/$(Bi_{1-x}Sb_x)_2Te_3$(2 nm)/$Bi_2Te_3$(2 nm)/$(Bi_{1-x}Sb_x)_2Te_3$(2 nm)/CGT(12 nm) heterostructures. (a) Temperature ($T$) dependence of the longitudinal resistivity ($\rho_{xx}$) on a logarithmic scale from 300 K to 2 K with variation of $x$. (b) $\rho_{xx}$ as functions of $T^{-1}$. The black broken line is the Arrhenius fitting result [$\rho_{xx} \propto \exp(T_0/T)$, where $k_B T_0 = 11$ meV] for the $x = 0.5$ sample. (c) Magnetic field ($\mu_0 H$) dependence of the Hall resistivity ($\rho_{yx}$) for the $x = 0.5$ sample at various $T$. (d) $T$ dependence of carrier density [electron: $n_{2D}$ (closed circles), hole: $p_{2D}$ (open circles)].

Figures S6(a) and 2(b) show the temperature ($T$) dependence of the longitudinal resistivity ($\rho_{xx}$) for the CGT(12 nm)/(Bi$_{1-x}$Sb$_x$)$_2$Te$_3$(2 nm)/Bi$_2$Te$_3$(2 nm)/(Bi$_{1-x}$Sb$_x$)$_2$Te$_3$(2 nm)/CGT(12 nm) samples shown in Fig. 3 of the main text; log$\rho_{xx}$ vs. (a) $T$ and (b) $1/T$. The samples with $x = 0.3$, 0.5 and 0.6, which exhibit large anomalous Hall angles, show double-step temperature dependence: increase in $\rho_{xx}$ with decreasing $T$ to 100 K, saturation at around 100 K, and increase again below 10 K. The $T$ dependence in the higher temperature region ($T > 100$ K) can be attributed to freezing of the bulk carriers in the BST/BT/BST layers. The curves at the higher temperature region in Fig. S6(b) are rounded and the fitting to the Arrhenius-type temperature dependence [$\rho_{xx} \propto \exp(T_0/T)$, where $T_0$ is the thermal activation energy] does not work well. The deviation from the Arrhenius-type $T$ dependence may suggest the presence of variable-range hopping transport among the localized states in the bulk part of the BST/BT/BST layer [S16-S18].

As $T$ is lowered below 100 K, $\rho_{xx}$ turns to be saturated. The saturation behavior ensures that the transport below 100 K is dominated by the surface states which are expected to show metallic $T$ dependence. However, below 10 K, $\rho_{xx}$ turns to increase again. This trend can be seen in the insulating samples. The $\rho_{xx}$ increase at low $T$ can be attributed to the reduction of surface carriers due to the exchange gap formation and/or their localization due to the disorder in the magnetized surface state. Such reduction of the surface carriers is also observed in Hall responses as shown in Fig. S6(c) for the $x = 0.5$ sample and in Fig. S6(d) for all the samples.

(ii) The attempt to estimate the size of the exchange gap

In the clean limit, the Hamiltonian of the TI surface state can be written as

$$H(k) = \hbar v_F(k_y \sigma_x - k_x \sigma_y) + \frac{\Delta}{2}\sigma_z, \tag{1}$$

where $v_F$ is the Fermi velocity, $\sigma_i$ ($i = x, y, z$) are Pauli matrices acting on spins, and $\Delta$ is the exchange gap. The Hall conductivity ($\sigma_{xy}$) is related to the Berry curvature, which increases as the Fermi level $E_F$ approaches $E = \pm\Delta/2$. When the Fermi level lies in the gap ($|E_F| < \Delta/2$), $\sigma_{xy}$ from two Dirac states (top and bottom surfaces) is quantized to $2 \times e^2/2h = e^2/h$. When the Fermi level lies above the gap ($|E_F| > \Delta/2$), $\sigma_{xy}$ can be obtained as [S19]

$$\sigma_{xy} = 2 \times \frac{e^2}{2h} \frac{\Delta/2}{\sqrt{(\hbar v_F)^2 k_F^2 + (\Delta/2)^2}} = \frac{e^2}{h}\frac{\Delta}{2|E_F|}. \tag{2}$$

We consider only the intrinsic contribution from the band structure and ignore other contributions from skew scatterings and side jumps; the latter two extrinsic mechanisms are unlikely to work in the present case judging from the features of low carrier density and high scattering rate contrary to the case of clean itinerant ferromagnets [S20]. In addition, the carrier density of the surface state ($n_{2D}$) is written as

$$n_{2D} = \frac{k_F^2}{2\pi} = \frac{E_F^2 - (\Delta/2)^2}{2\pi(\hbar v_F)^2}, \tag{3}$$

therefore it is possible to estimate $E_F$ and $\Delta$ from the measured $\sigma_{xy}$ and $n_{2D}$.

Although we are already aware that our samples are influenced by disorder/inhomogeneity, it is inviting to estimate rough energy scales of $E_F$ and $\Delta$ by putting the experimentally measured values in Eqs. (2) and (3). We focus on the data of the bulk-insulating CGT(12 nm)/(Bi$_{1-x}$Sb$_x$)$_2$Te$_3$(2 nm)/Bi$_2$Te$_3$(2 nm)/(Bi$_{1-x}$Sb$_x$)$_2$Te$_3$(2 nm)/CGT(12 nm) sample ($x = 0.5$) at 2 K. From the normal Hall coefficient $R_H$ at high magnetic field, we estimate $n_{2D} = 1.9 \times 10^{12}$ cm$^{-2}$ assuming that all the contributions to $R_H$ are from the surface state. The value of anomalous Hall conductivity is $\sigma_{xy} = 0.16 e^2/h$ as shown in Fig. 3 of the main text. Putting these values together with $v_F = 3.6 \times 10^5$ m s$^{-1}$ from the ARPES result [S8], $E_F$ and $\Delta$ are estimated to be 83 meV and 27 meV, respectively.

The estimated values may be valid if the amplitude of the potential fluctuation due to inhomogeneity is much smaller than $E_F$ and $\Delta$. However, as is the case in pristine and magnetically-doped TIs, the potential fluctuation can amount to be several tens meV [S16, S17]. Therefore, we should be cautious in applying the clean limit model to analyze the experimental result. In the presence of inhomogeneity, $E_F$ and $\Delta$ can have spatial variation. Then the sample becomes a patched network of different $E_F$ and $\Delta$. In such a case, the measured $\sigma_{xy}$ becomes a complicated combination of local $\sigma_{xy}$. Nevertheless, the estimated $E_F$ value is well below the bulk bandgap value (~ 300 meV) and the estimated $\Delta$ value is large as ~ 1/3 of the $E_F$. This may point to the necessary existence of the exchange gap, albeit spatially varying, so as to give rise to such a large anomalous Hall conductivity (or Hall angle) as observed. For more quantitative discussion, we need to estimate the amplitude of the potential fluctuation in the CGT/BST system and to construct an elaborate model taking the potential fluctuation into account, which will be an issue of future work.

## VII. Characterization and magneto-transport data of BST/CGT bilayer structures.

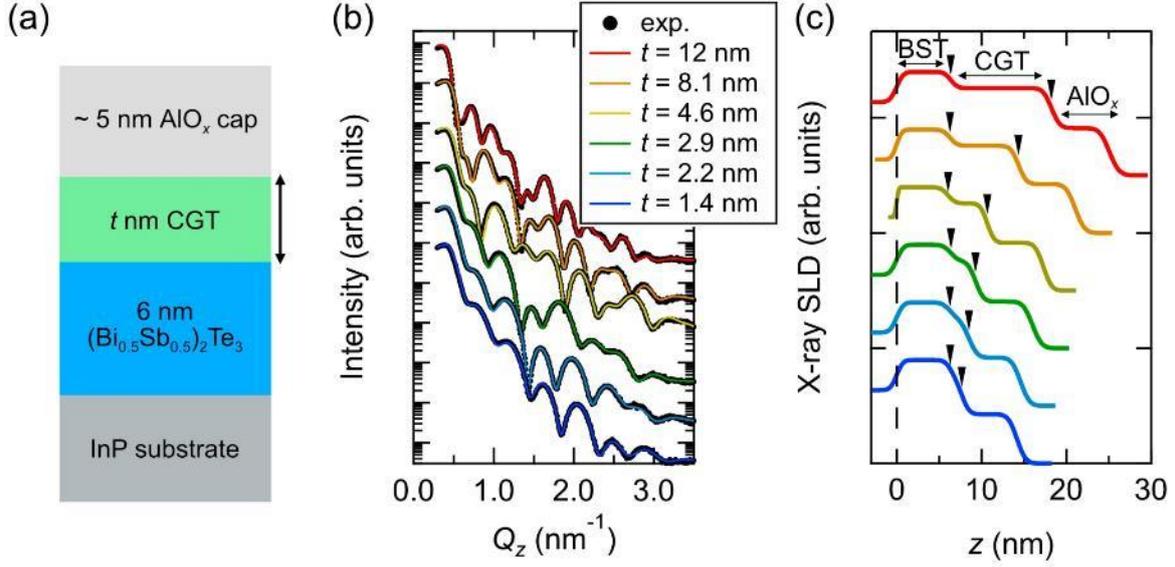

FIG. S7. X-ray reflectivity (XRR) for BST(6 nm)/CGT($t$ nm)/AlO$_x$(5 nm) structures. (a) Schematic layout of the sample structure. (b) Measured (dots) and fitted (solid lines) XRR curves on a logarithmic scale for the BST/CGT bilayer structures with the various CGT thicknesses ($t$) as a function of momentum transfer ($Q_z$). (c) X-ray SLDs as a function of the distance from the InP substrate surface, $z$. Triangles indicate the bottom and the top surface of CGT layers.

By measuring x-ray reflectivity (XRR), we investigated CGT thickness ($t$) of the BST/CGT bilayer structures [Fig. S7(a)] used in Fig. 4 of the main text. Figure S7(b) and S7(c) show the XRR reflectivity curves and the depth profiles of the x-ray SLD, respectively. The SLD profiles, representing the thicknesses of layers and sharp interfaces, were simulated by the same procedure as the analysis of the data in Fig. 2(c) of the main text. We note that the root-mean-square roughness of CGT is as small as 0.74(5) nm.

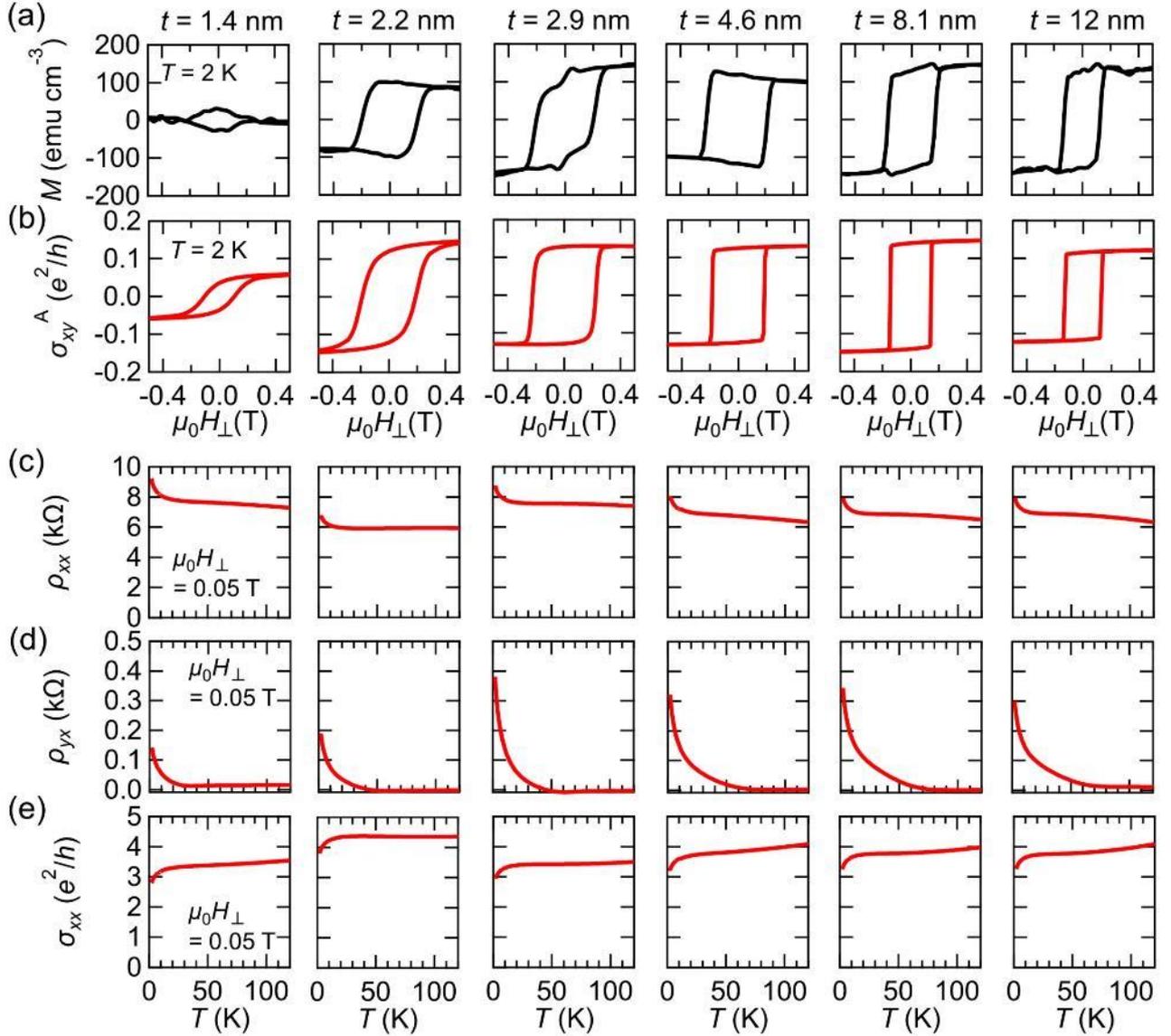

FIG. S8. Additional magneto-transport data of BST(6 nm)/CGT($t$ nm)/AlO$_x$(5 nm) structures. (a, b) Out-of-plane magnetic field dependence ($\mu_0 H_\perp$) of the magnetization ($M$) (a) and the anomalous Hall conductivity ($\sigma_{xy}^A$) (b) at 2 K in the BST/CGT ($t$ = 1.4, 2.2, 2.9, 4.6, 8.1, 12 nm) structures. (c-e) Temperature ($T$) dependence of the longitudinal sheet resistivity ($\rho_{xx}$) (c), the Hall resistivity ($\rho_{yx}$) (d), and the longitudinal sheet conductivity ($\sigma_{xx}$) (e) measured under field cooling at $\mu_0 H_\perp$ = 0.05 T.

Figures S8(a) and S8(b) show the out-of-plane field ($H_\perp$) dependence of the magnetization ($M$) and the anomalous Hall conductivity ($\sigma_{xy}^A$) derived by subtracting the $H_\perp$-linear component at $T$ = 2 K, respectively. Both magnetic hysteresis loops in $M$ and $\sigma_{xy}^A$ are consistent with each other. We note that the negative slopes of the magnetization for $t = 1.4$ and 2.2 nm originate from the subtraction procedure of diamagnetism of InP substrates, which is conducted by a reference measurement of an InP substrate without films. Therefore, in Fig. 4(d) of the main text, we used the zero-field value of $\sigma_{xy}^A$ and $M$ to eliminate the residual substrate contribution. In Fig. S8(c)-S8(e), the temperature ($T$) dependence of the longitudinal sheet resistivity ($\rho_{xx}$), the Hall resistivity ($\rho_{yx}$) and the longitudinal sheet conductivity ($\sigma_{xx}$) are shown for supporting the data of Fig. 4(d) in the main text.